\documentstyle[prl,aps,floats,psfig]{revtex}

\begin{document}
\author{ Benjamin D.\ Wandelt$^1$, Eric Hivon$^1$, Krzysztof M.\ G\'orski$^{1,2}$}
\address{ $^1$Theoretical Astrophysics Center,
	Juliane Maries Vej 31,
	DK-2100 Copenhagen,
	Denmark\\
	$^2$Warsaw University Observatory, Warsaw, Poland}
\twocolumn[\hsize\textwidth\columnwidth\hsize\csname@twocolumnfalse\endcsname

\title{Cosmic microwave background anisotropy power spectrum statistics for high precision cosmology}

\maketitle
\begin{abstract}
As the era of high precision cosmology approaches,
the empirically determined
power spectrum of the microwave background anisotropy, 
$C_l$, will provide a crucial test for cosmological theories. 
We present a unified semi-analytic framework 
for the study of the statistical properties of the $C_l$ coefficients
computed from the results of balloon, ground based, and
satellite experiments.
An illustrative application shows that
commonly used approximations {\it bias} the estimation of the
baryon parameter $\Omega_b$ at the 1\% level even for a satellite 
capturing as much as
$\sim 70$\% of the sky.
\end{abstract}
\date{\today}
]

During the next few years ground based observations and
balloon missions\cite{nonsatellite} as well as
satellite observations\cite{map} promise exquisite
determinations of the 
power--spectrum, $C_l$, of cosmic microwave background (CMB) anisotropies. 
Recently, pioneering work\cite{knox,jungman,BET,ZSS,BJK} 
has shown, in the context of inflationary cosmogonies,
the tremendous impact these missions will have on our knowledge of
cosmological parameters. It is therefore of paramount
importance to study the statistical properties of these
quantities in detail. 
The fact that the full solution of the associated 
likelihood problem
poses severe computational difficulties, has prompted many practical 
analyses
to invoke one or more of the following simplifying assumptions: 1) the
observed $C_l$ are approximately independent due to nearly full and
uniform sky
coverage, 2) their sampling distributions do not change appreciably
from the  $\chi^2$ distributions which apply for the full sky, apart
from a rescaling of the mean by the sky fraction to account
for lost power, and 3) the sampling distributions are well-approximated
by a Gaussian which has the same first and second moments as these
rescaled $\chi^2$ distributions. 

There are
several reasons to assess these approximations and, if necessary, go
beyond them. 
One reason is that in the short term 
balloon and ground based experiments
will provide the leading edge science results to the field. Due to
practical limitations we cannot hope to even come close
to full sky coverage with these types of experiments. The observation
regions are often ring--shaped or cover a circular region of the sky
which subtends a small solid angle.  It is an urgent matter to assess
statistically how these imminent experiments can constrain the power
spectrum and cosmological parameters.

Further, the planned satellite missions are aiming to
determine the $C_l$ to sub--percentage accuracy. 
In the case of the Planck Surveyor mission,
this has given us the hope
of detecting small effects such as 
secondary anisotropies which are due to nonlinear
gravitational effects on the CMB photons during the free--streaming
epoch. This 
is an important issue because a detection would break otherwise
present parameter
degeneracies and allow a consistent parameter estimation from CMB data
alone\cite{SE}.  

More generally, 
the impact CMB observations will have on cosmology makes it important
to use  approximations 
in a controlled way. For example, in the
analysis of COBE--DMR data it was realized that 
using Gaussian approximations for
quadratic quantities introduced systematic biases\cite{banday}.
At the same time, we need approximations to make feasible the analysis of
the huge CMB data sets we expect in the coming years.

To provide a quantitative basis for this discussion  we
present in this Letter a semi-analytic 
framework for the calculation of the sample statistics
of the $C_l$ for theories which generate a Gaussian CMB sky. 
This framework is exact for survey geometries and noise patterns
which obey rotational symmetry about one, 
arbitrary axis, such as polar cap shaped regions, Galactic cuts,
rings, and annuli of any size and at any latitude.
We find that our methods also allow dealing with arbitrary noise patterns
to very good accuracy and conclude by illustrating their use in a first
application.

The full sky of CMB temperature fluctuations can
be expanded in spherical harmonics, $Y_{lm}$,
as\footnote{We assume that there is insignificant signal power in modes
with $l>l_{max}$ and use the convention that sums over $m$ run from
$-l_{max}$ to $l_{max}$ and all quantities with index ${lm}$ vanish
for $m>l$.}
\begin{equation}\label{alms}
T({\bf \gamma})=\sum_{l=0}^{l_{max}}\sum_{m}a_{lm}Y_{lm}(\gamma)
\end{equation}
where $\mathbf{\gamma}$ denotes a unit vector pointing at polar angle $\theta$ and
azimuth $\phi$.
A Gaussian cosmological theory states that the $a_{lm}$ are Gaussian
distributed with zero mean and specified variance 
$C^{theory}_l\equiv\langle\left\vert {a_{lm}}\right\vert ^2\rangle$. 
Hence, for noiseless, full sky measurements,
each measured $C_l$ independently follows a
$\chi^2$--distribution with $2l+1$ degrees of freedom and mean $C^{theory}$.

Owing to Galactic foregrounds, limited surveying time or other
constraints inherent in the experimental setup, the
temperature map that comes out of an actual measurement will be
incomplete. In addition, a given scanning strategy will produce a
noise template. We model the noise as a Gaussian field $T_N$
with zero mean which is independent from pixel to pixel and modulated
by a spatially varying rms amplitude $W_N({\mathbf{\gamma}})$. Therefore the
observed temperature 
anisotropy map is in fact
\begin{equation}
\tilde{T}({\mathbf{\gamma}})=W({\mathbf{\gamma}})\left[T({\mathbf{\gamma}})+W_N({\mathbf{\gamma}})T_N({\mathbf{\gamma}})\right]
\label{realtemp}
\end{equation}
where $W$ is unity in the observed region and zero elsewhere.

Expanding $\tilde{T}$ as in 
Eq.\ (\ref{alms}) produces a set of {\em correlated} Gaussian
variates $\tilde{a}_{lm}$ for the signal and $\tilde{a}_{N\;lm}$ for the
noise. These combine into
power spectrum coefficients
\begin{equation}
\tilde{C}_l=\frac1{2l+1}\sum_m
\left\vert{\tilde{a}_{lm}+\tilde{a}_{N\;lm}}\right\vert ^2
\label{pcldef}
\end{equation}
whose statistical properties differ from the ones of the $C_l$. 
We therefore refer to these quantities as {\it pseudo}-$C_l$. 
In what follows we will discuss the statistical
properties of these quantities, restricting ourselves to a
presentation of results. We relegate detailed derivations and
implementational issues to a future publication\cite{future}.

\label{sec:pclstats}

If we assume white noise, $C^N\equiv C^N_l$,
then each term in the sum Eq.\ (\ref{pcldef}) has expectation value 
\begin{equation}
\sigma^2_{lm}=\frac{\sum_{l'm'}C_{l'}\left\vert{W_{l'm'\;lm}}\right\vert
	^2+
	C^N\!\!\int_O d{\mathbf{\gamma}} W_N({\mathbf{\gamma}})^2\lambda^2_{lm}(\theta )}{2l+1}
\label{slm}
\end{equation} 
where we abbreviate the polar part of the $Y_{lm}$ as
$\lambda_{lm}\equiv \sqrt{\frac{2l+1}{4\pi}\frac{(l-m)!}{(l+m)!}}P_l^m$,
the $W_{l'm'\;lm}$ is the matrix element of $W$ in Eq.\ (\ref{realtemp}) 
in a spherical harmonic basis 
and $\int_O$ integrates over the observed region of the sky. Note that
the cross-term which comes from expanding the square in Eq.\ (\ref{pcldef})
has vanishing expectation value because signal and noise are assumed
to be uncorrelated.

To elucidate the correlation structure between the $\tilde{a}_{lm}$ we
omit the $a_{N\;lm}$ for simplicity and write 
Eq.\ (\ref{pcldef}) as
a quadratic form in the independent normal variates
$\alpha_{lm}=a_{lm}/\sqrt{C_l}$. This  
defines the coupling matrix ${\cal{M}}^{(l)}$ such that 
$\tilde{C}_l=\alpha^\ast {\cal M }^{(l)} \alpha$. For the full sky
${\cal{M}}^{(l)}_{l_1m_1\;l_2m_2}=
   \frac{C_l}{(2l+1)}\sum_m\delta_{ll_1}\delta_{mm_1}\delta_{ll_2}\delta_{mm_2}$.
In the case of azimuthal symmetry ${\cal{M}}^{(l)}$ is block diagonal:
\begin{equation}
{\cal{M}}^{(l)}=\frac1{2l+1} \left\{\bigoplus_m\left(V^{lm}\otimes V^{lm}\right)\right\}
\label{correl}
\end{equation}
where $V_{l'}^{lm}=W_{l'm\;lm}\sqrt{C_{l'}}$. 
Each block is the Cartesian product of $V$ with itself. Moreover, if
$W_N$ has the same azimuthal symmetry as $W$, 
this argument can be extended to include the noise.
Eq. \ref{correl} is the key fact which allows the derivation and cheap
evaluation of the exact results we obtain.
(To compute the formulas we present in this Letter, 
generating the $W$
matrices is the most costly operation. For a
maximum $l$ of 1024 this takes just over 
1 minute on a single R10000 CPU.)

It follows that while the $\tilde{a}_{lm}+\tilde{a}_{N\;lm}$ terms in
Eq.\ \ref{pcldef}
are correlated for
different $l$, they will be
uncorrelated for different $m$. As a consequence, we can view the
$\tilde{C}_l$ as sums of independent 
Chi-squared ($\chi^2$) variates, each with one degree of freedom but
different expectations. This allows us to 
solve for their statistical properties exactly. 

\label{quad:moments}
We can compute the distribution of these sums
analytically using the method of characteristic
functions. Defining
$s^{(l)}_m=(2/\sigma^2_{lm})$, we
produce a closed form solution in terms of 
incomplete gamma functions $\gamma(\alpha,x)$:
\begin{equation}
\!\!\!\!\!\!\!\!P(\tilde{C}_l)=A^{(l)}
\sum_{m=1}^{l}\frac{e^{-s^{(l)}_m \tilde{C}_l}\,\gamma\!\left(\frac{1}{2},(s^{(l)}_0-s^{(l)}_m)\tilde{C}_l\right)}
{\sqrt{(s^{(l)}_0-s^{(l)}_m)}
\prod_{\;m=1}^{'\;l} (s^{(l)}_{m'}-s^{(l)}_{m})}
\label{probdist}
\end{equation}
where the primed product symbol $\prod^{'}$ only multiplies
factors 
which have $m\neq m'$ and the normalization constant is
$A^{(l)}\equiv\frac{\sqrt{s^{(l)}_0}\prod_{m=1}^{l}s^{(l)}_m}{2
\Gamma(\frac{3}{2})}$. 

We can then obtain the cumulants of the 
pseudo-$C_l$ as $\kappa_n=2^{n-1}(n-1)!\sum_m(\sigma_{lm}^2)^{n}$. Any moment
can be written in terms of these cumulants.
We give the following expressions
for the mean, variance, skewness 
$\beta_1=\frac{\langle(\Delta\tilde{C}_l)^3\rangle}{\langle(\Delta\tilde{C}_l)^2\rangle^{\frac32}}$
and kurtosis
$\beta_2=\frac{\langle(\Delta\tilde{C}_l)^4\rangle}{\langle(\Delta\tilde{C}_l)^2\rangle^2}$ as examples:
\begin{equation}
\begin{tabular}{cc}
$\langle\tilde{C}_l\rangle=\sum_m \sigma^2_{lm}$,& 
   $\langle(\Delta\tilde{C}_l)^2\rangle=2\sum_m\sigma^4_{lm}$\\
\\
$\beta_1=2^{\frac32}\frac{\sum_m{\sigma_{lm}^6}}{\left(\sum_m{\sigma_{lm}^4}\right)^{\frac32}},$&
$\beta_2={12}\frac{\sum_m{\sigma_{lm}^8}}{\left(\sum_m{\sigma_{lm}^4}\right)^2}$.\\
\end{tabular}
\label{meanetc}
\end{equation}

\label{joint}
The covariance between the pseudo-$C_l$ can be written as 
$\left\langle{\Delta \tilde{C}_l \Delta C_{l'}}\right\rangle={\rm tr}\, {{\cal{M}}^{(l)}{\cal{M}}^{(l')}}$ which
reduces to 
\begin{equation}
\left\langle{\Delta \tilde{C}_l \Delta C_{l'}}\right\rangle=\sum_m\left(\sum_{l_1} V^{lm}_{l_1}V^{l'm}_{l_1}\right)^2
\label{covmat}
\end{equation}
in the azimuthally symmetric case.
All higher order joint moments are computed similarly in terms of traces
of products of ${\cal{M}}^{(l)}$ of various $l$\cite{future}.

\label{approx}
There are some important situations where the noise pattern does {\em not}
follow the azimuthal symmetry of the survey geometry. In the case of
the Planck satellite the scanning strategy is approximately centered
on the ecliptic poles, while the Galactic cut is 
tilted through $\approx 60^\circ$
with respect to this. In this case Eq.\ (\ref{probdist}) becomes an
approximation. 
We found it to be very accurate indeed
to continue using these distributions with the
$\sigma^2_{lm}$ computed for an asymmetric $W_N$, even for a strongly
asymmetric noise pattern. This approximation will be worst in the
least interesting, noise dominated regime at very high $l$.
Note that the 
$\sigma^2_{lm}$ and hence the $\langle \tilde{C}_l \rangle$ 
remain exact (because the $\lambda_{lm}$ are independent of
the azimuth) but the remarks leading to Eq.\ (\ref{probdist}) are no longer
exactly true. For applications
the final justification comes from the excellent agreement we find
when we check against our Monte Carlo simulations.

\label{quad:mc}%

To test our results we performed 3328 Monte Carlo (MC) 
simulations for a high resolution CMB satellite, such as MAP or
Planck (resulting in a sky coverage comparable to COBE). We simulated
realizations of the CMB sky in the standard cold 
dark matter model($\Omega_m=1$, $\Omega_bh^2=0.015$,
$H_0=70$km/s/Mpc). From 
these maps we carved out a $\pm 20^\circ$ Galactic cut and contaminated the
remaining area with spatially modulated Gaussian white noise of
maximum rms temperature $124\mu K$ per pixel of characteristic size
3.4 arcminutes. We use
a tilted noise template $W_N=\sqrt{\sin\theta_E}$, where $\theta_E$ is the
ecliptic latitude, as a simple 
model of the noise pattern which
would result from scanning along meridians through the
ecliptic poles. We then Fourier analyzed these
maps and stored the resulting $\tilde{C}_l$. A Kolmogorov--Smirnov test failed
to detect deviations between the distributions of this
MC population and Eq.\ (\ref{probdist}) at 99\% confidence, which
validates our semi--analytical expressions. 

\begin{figure}[tf]
\centerline{\psfig{file=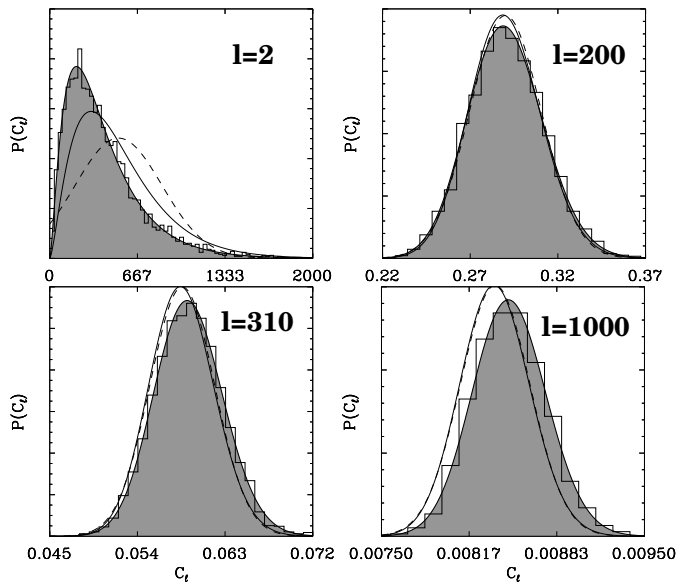,width=3.375in}}
\caption{The Pseudo-$C_l$ distributions, Eq.\ \protect\ref{probdist}, 
for the  standard cold dark matter model (shaded), 
the $\chi^2$ (solid lines) and the Gaussian (dashed) approximations
compared to Monte 
Carlo simulations (histograms) for
l=2,200,310,1000.}
\label{pcldists}
\end{figure}

To give a visual impression of
the resulting probability densities we show four of them 
in Figure \ref{pcldists} together with the results from the MC
simulations. 
Also shown in this Figure are the $\chi^2$
distributions which the $C_l$ would follow in the full sky case as
well as the commonly used Gaussian approximation.
These are mean adjusted to account for the lost solid angle due
to the Galactic cut. At $l\lesssim 30$ the difference is
striking. For higher $l$ the Gaussian approximation becomes better as
higher moments die away by dint of the Central Limit Theorem, but
there remain visible systematic differences to the true distributions.
In particular, there is a residual shift in the mean and the
approximations tend to be slightly narrower than the histograms for
very high $l\approx 1000$.

\begin{figure}[tf]
\centerline{\psfig{file=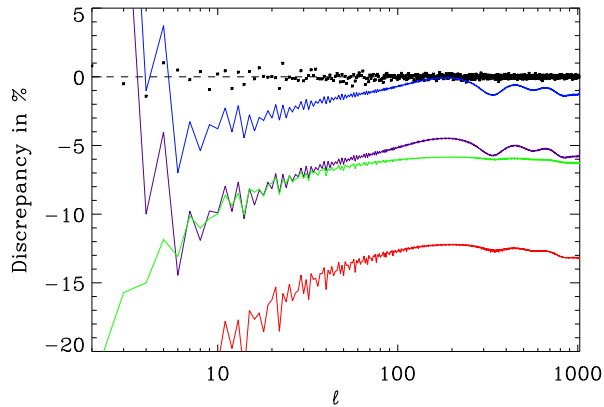,width=3.375in}}
\caption{The solid lines are, top to bottom, 
the percentage discrepancy between 
the mean, standard deviation, skewness, and kurtosis of the 
$\chi^2$ approximation
and the Pseudo-$C_l$ distributions. The stars are the 
$\langle \tilde{C}_l \rangle$ computed
from 3328 Monte Carlo runs, showing excellent agreement only
limited by Monte Carlo noise to better than 0.1 \% for all $l$.}
\label{moments}
\end{figure}

This becomes a more quantitative observation when looking at
the percentage discrepancies between the mean and variances as a function of
$l$ in Figure \ref{moments}. 
The discrepancy is  of the
order of 1 \% in the mean and 5\% in the standard deviation 
on most scales, except for $l<70$ where the effect is 
larger. These discrepancies are important at the level of
precision of future almost full sky missions. For medium and small
sky coverage the mode couplings are stronger and we expect this to
have an even larger effect on the probability distributions. 
We also compare the 
skewness $\beta_1$ and kurtosis 
$\beta_2$
of the $\chi^2$
distributions to our distributions. The percentage difference is
larger than for the first two moments but arguably less important 
at large $l$, since $\beta_1$ and $\beta_2$ decay as
$(2l+1)^{-\frac{1}{2}}$ and $(2l+1)^{-1}$, respectively. 

\begin{figure}[tf]
\centerline{\psfig{file=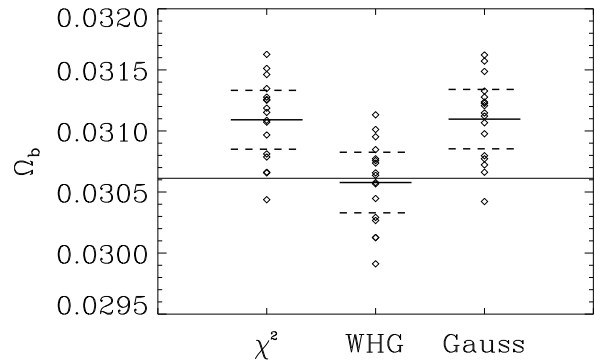,width=3.375in}}
\caption{One parameter likelihood estimates of $\Omega_b$. Shown are the
means (solid) 
within $\pm 3$ standard errors in the mean (dashed) 
of estimations from 17 realizations (diamonds) of
the sky. Both the $\chi^2$ (left)
and the Gaussian (right) approximations produce a bias, 
while our approximation (center) has no detectable bias. }
\label{obbias}
\end{figure}

\label{parest}
As a first application we study the effect of
approximating the likelihood for parameter estimation. 
Since our distributions have the correct means, we simply
multiply them together 
for a simple, unbiased approximation to the likelihood. This is
conservative since the marginal distributions have all correlations
{\em integrated} out and we will therefore overestimate the
error bars on the $C_l$. Using this likelihood, as well as the
Gaussian and $\chi^2$ approximations, we attempt to estimate
the baryon parameter $\Omega_b$ (holding all other parameters
constant) from several randomly selected
realizations in our MC pool. The results are shown in
Figure \ref{obbias}. As expected, Gauss and $\chi^2$  consistently
find estimates which are biased about 1.6\% high, 
3 standard errors of the mean away from the true value, 
while our likelihood gives
a perfect fit. Since the moment 
discrepancies depend on the underlying cosmological theory, this level
of bias is only indicative of the general level of error introduced by
using the Gaussian or $\chi^2$ approximations.

We note that  numerical techniques have recently been developed
\cite{OhSpergelHinshaw} which allow the computational
solution of the power spectrum estimation for high $l$ applications. 
These methods are efficient for the case of almost
full sky observations.
A purely numerical approach
still requires significant computational resources, especially if
there is signal in modes with $l>1000$. The results
presented in this Letter should
be seen as complementary to such calculations. An analytical framework 
admits a more fundamental approach 
to understanding and is a useful yardstick against which numerical
work can be tested or from which 
approximate methods can be derived.

To illustrate, we suggest the following computationally cheap and accurate
approximation recipe, motivated by the fact that higher
moments of the $\tilde{C}_l$ distributions die away quickly with higher $l$:
on measuring the $\tilde{C}_l$ from the sky, fit a smooth curve through
them and use this fit as the input theory for our framework. Then calculate
the corrections to the sample means and
variances using Eqs.\ (\ref{slm},\ref{meanetc}).
These can in turn be used for simple $\chi^2$ fitting in the usual
way, at least for high $l\gtrsim 100$. 
Since the discrepancies are small for large sky coverage this
will produce a fit which is accurate to second order in the
discrepancy,  safely within the regime of accuracy envisaged for
modern satellite missions.

\label{quad:conclusions}

Apart from the obvious applications to experiments with small and
medium sky coverage such as balloons or ground--based missions,
many further uses of this framework are conceivable. For example, one could
1) extend this treatment to multi--parameter fits,
2) design 'optimal' scanning strategies, encoded in $W_N$, 3) use
Eq.\ (\ref{probdist})  once a theory is estimated to check $l$ by $l$
for consistency with the assumption of Gaussian primordial
fluctuations and 4) assess more realistically if secondary
anisotropies will be detectable with future CMB missions. 
Finally, all ingredients are there to refine the approximation to
the joint likelihood we used in this Letter by taking into account the
covariances 
Eq.\ (\ref{covmat}), for example by using a multivariate Edgeworth expansion
around the peak of the likelihood.


To summarize, we have presented a unified theoretical framework for the study
of power spectrum determinations of balloon, ground based and
satellite experiments. We go beyond current approximations and present
a semi-analytic formalism for the computation of sampling
distributions of the $C_l$ for any Gaussian cosmological model and a
large and important class of surveying strategies. We show that
applying this method to the estimation of $\Omega_b$ from simulated data
is unbiased and hence superior to commonly used analytic approximations which
bias the result at the percent level. A number of applications 
and extensions of this formalism remain to
be explored\cite{future}.

We wish to thank A.\ J.\ Banday for stimulating discussions. This research
was supported by the Dansk Grundforskningsfond through its funding for
TAC. 




%
%

\end{document}